\documentclass[preprint,showpacs,preprintnumbers,amsmath,amssymb]{revtex4-1}
\usepackage{amssymb}
\usepackage{amsmath}
\usepackage{graphicx}
\usepackage{epsfig}
\usepackage{subfigure}
\usepackage{amsfonts}

\begin{document}

	\title{Implementing quantum walks with a single qubit}
	
	\author{Qi-Ping Su$^1$}
	\author{Shi-Chao Wang$^1$}
	\author{Yan Chi$^1$}
	\author{Yong-Nan Sun$^1$}
	\author{Li Yu$^1$}
	\author{Zhe Sun$^1$}
	\author{Franco Nori$^{2,3,4}$}
	\email{fnori@riken.jp}
	\author{Chui-Ping Yang$^{1,5}$}
	\email{yangcp@hznu.edu.cn}
	\address{$^1$Department of Physics, Hangzhou Normal University, Hangzhou 311121, China}
	\address{$^2$Theoretical Quantum Physics Laboratory, RIKEN Cluster for Pioneering Research, Wako-shi, Saitama 351-0198, Japan}
	\address{$^3$RIKEN Center for Quantum Computing (RQC), Wako-shi, Saitama 351-0198, Japan}
	\address{$^4$Department of Physics, University of Michigan, Ann Arbor, Michigan 48109-1040, USA}
	\address{$^5$Quantum Information Research Center, Shangrao Normal University, Shangrao 334001, China}

\begin{abstract}
Quantum walks have wide applications in quantum information, such as universal quantum computation, so it is important to explore properties of quantum walks thoroughly. We propose a novel method to implement discrete-time quantum walks (DTQWs) using only a single qubit, in which 
both coin and walker are encoded in the two-dimensional state space of a single qubit, 
operations are realized using single-qubit gates only,
and high-dimensional final states of DTQWs can be obtained naturally.  
With this ``one-qubit'' approach, DTQW experiments can be realized much more easily, compared with previous methods, in most quantum systems and all properties based on quantum states of DTQWs (such as quantum correlation and coherence) can be investigated.
By this approach, we experimentally implement one-particle and two-particle DTQWs with seven steps using single photons. Furthermore, we systematically investigate quantum correlations and coherence (based on the full state of the coin and walker) of the DTQW systems with different initial states of the coin, which have not been obtained and studied in DTQW experiments. 
As an application, we also study the assisted distillation of quantum coherence using the full state of the two-particle DTQW from the experiment. The maximal increase in distillable coherence for high-dimensional mixed states is investigated for the first time by obtaining its upper and lower bounds.
Our work opens a new door to implement DTQW experiments and to better explore  properties of quantum walks.
\end{abstract}



\maketitle

\section{Introduction}
Quantum walks can be divided into discrete-time quantum walk (DTQW) and contunuous-time quantum walk (CTQW). In general, the CTQW is easier to implement in experiments, while the DTQW is faster and has more degrees of freedom \cite{q0}.
DTQW has a wide range of applications in quantum algorithms \cite{q1,q2,q3}, universal quantum computation \cite{q4,q5,q6}, quantum
simulations \cite{q601,q7,q701}, topological phases \cite{q8,q9}, state preparation and transfer, and other areas \cite{q901,q10,q11,q12}. 
Moreover, the experimental implementation of DTQW has been achieved in several quantum
systems, such as linear optics \cite{q3,q7,q9,q130,q13,q14,q141}, ion
traps \cite{q15,q16}, neutral atom traps \cite{q17}, and circuit QED \cite{q18}.
In DTQW experiments, the probability distribution of the walker and the density matrix of the low-dimensional coin (after taking partial trace over the walker) are usually adopted to study the properties of DTQWs, which are easy to measure.
A few correlation functions and the entanglement between the coin and the walker, based on one of the two measurable quantities, have also been adopted to reveal properties of DTQWs (e.g., see Ref. \cite{q7, q141, q19, q20,q290,q292,q293}).

However, after a review of the literature, we find that for DTQWs with multiple steps, quantum quantities based on the full state of the walker and the coin, such as quantum correlations and quantum coherence, have not been used to investigate DTQWs in experiments.
This is mainly because the high-dimensional quantum states (i.e., joint states of the walker and the coin in multi-step DTQWs) are very difficult to obtain in DTQW experiments.
Note that the joint state of the walker and the coin, which contains all information, is a key element to thoroughly study general properties of DTQWs.
Researches in this area can help to explore deep properties of quantum walks and expand their applications.

In this work, we first propose a novel method for the implementation of DTQWs, in which both the coin and walker are encoded by a single qubit. 
With this one-qubit method, the whole final states of DTQWs after many steps can be  naturally obtained in principle, which can be used to study deep properties of DTQW systems.
Because only a single qubit is needed, the resource required for implementing DTQWs is greatly reduced and the experimental implementation of DTQWs becomes extremely simple.
It enables one to implement multi-particle DTQWs efficiently, which can provide an additional computational power and can be used to improve simulation performances in complex tasks \cite{q6,q294}.
Furthermore, the implementation of DTQW in one qubit can be adopted to realize DTQWs in most quantum systems. 

By this one-qubit method, we experimentally implement one-particle and two-particle DTQWs with seven steps in linear optics. We obtain all final joint states of the walker and the coin in the DTQW experiment and systematically study quantum correlations and coherence of the DTQW systems with different initial states of the coin. 
The experimental results fit well with the theoretical results.
As an application, we also investigate the assisted distillation of quantum coherence for high-dimensional pure states and mixed states using the full states of the two-particle DTQW from the experiment. 


\section{Implementing DTQW with a single qubit}

In a standard DTQW, a walker moves with respect to the state of a coin.
The evolution of the walker and the coin can be characterized by an
operator $U=T\cdot S$, where $T$ is the shift operator of the walker and $S$ is the coin operator. For example, in each step of a one-dimensional (1D)
DTQW, the coin, with states $|0\rangle_c$ and $|1\rangle_c$, is tossed by   
\begin{equation}
	S=I_w\otimes\left( \cos{\theta}|0\rangle_c\langle0|-\sin{\theta}|0\rangle_c\langle1|+\sin{\theta}|1\rangle_c\langle0|+\cos{\theta}|1\rangle_c\langle1| \right),
	\label{coin}
\end{equation}
where $\theta\in[0,\pi)$, and $I_w$ is an identity operator of the walker. Then, the walker is shifted by
\begin{equation}
	T=\sum_n|n-1\rangle\langle
	n|\otimes|0\rangle_c\langle0|+|n+1\rangle\langle
	n|\otimes|1\rangle_c\langle1|, \label{walker}
\end{equation}
where the integer $n$ represents the position of the walker in a 1D line.

Now we introduce how to implement a 1D DTQW using one qubit only. Both coin and walker of the DTQW are encoded in the single qubit.
The coin is encoded by two bases $|0\rangle_q$ and $|1\rangle_q$ of the qubit, and the position of the walker is encoded by the phase of the two bases. Thus, a general state of a 1D DTQW system is encoded as 
\begin{equation}
\sum_n(a_n|n\rangle|0\rangle_c+b_n|n\rangle|1\rangle_c)\rightarrow \frac{1}{N}\sum_n(a_ne^{in\phi}|0\rangle_q+b_ne^{in\phi}|1\rangle_q),\label{A1}
\end{equation}
where $\sum_n(|a_n|^2+|b_n|^2)=1$ and $N$ is the normalized coefficient. Though this encoding is invalid for all $a_n=0$ or all $b_n=0$, in most cases this encoding is valid.

\begin{figure*}[t]
	\begin{center}
		\includegraphics[width=12 cm, clip]{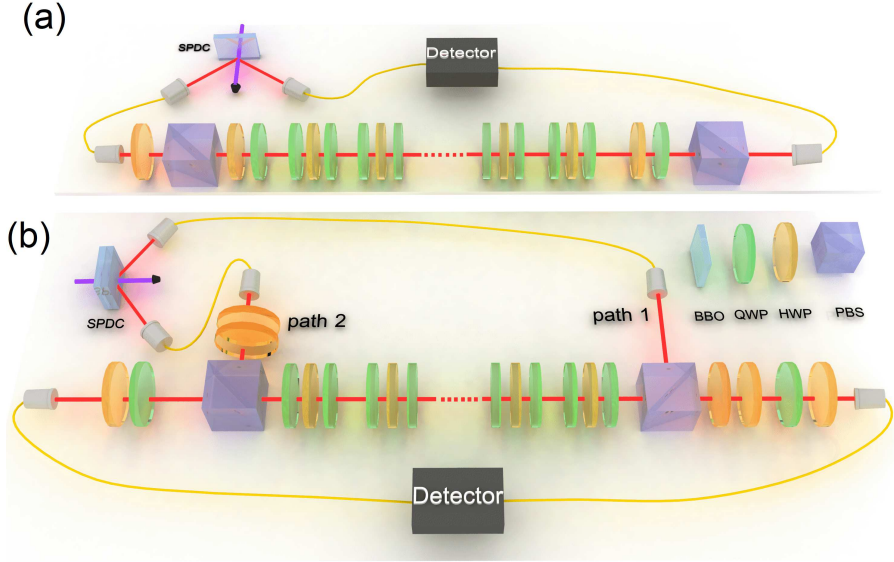}
		\caption{(a) Schematic of the experimental setup for the one-particle DTQW in linear optics. (b) Schematic of the experimental setup for the two-particle DTQW. Photon pairs are injected into path 1 and path 2, respectively. BS: beam splitter. HWP: half-wave plate. PBS: polarizing beam splitter. QWP: quarter-wave plate.}\label{fig1}
	\end{center}
\end{figure*}

Now we show how to readout the information $a_n$ and $b_n$ from the encoded qubit. The above encoding should be repeated ($2M-1$) times, where $M$ is the total number of $a_n$ (equals to the total number of $b_n$). 
For the $k^{th}$ encoding, the phase $\phi$ involved in Eq.~(\ref{A1}) is replaced by $\phi_k$ ($k=1,2,...,2M-1$). In each encoding, the state of the qubit can be measured, and the values of the ratio
$r_k=(\sum_na_ne^{in\phi_k})/(\sum_nb_ne^{in\phi_k})$
can be obtained from experiments. In this way, one obtains ($2M-1$) equations for $a_n$ and $b_n$ as 
\begin{equation}
	\sum_n(a_ne^{in\phi_k}-r_kb_ne^{in\phi_k})=0.\vspace*{-0.05in}\label{e}
\end{equation} 
We need to solve these equations, with the normalized condition: $\sum_n(|a_n|^2+|b_n|^2)=1$. Then, all $a_n$ and $b_n$ are obtained and thus the whole quantum state is readout. 

By this implementation method, the circuit and operations in DTQW experiments are extremely simple, because only one physical qubit is required.
The implementing can be sped up by $m$ times if $m$ qubits are adopted to run the same DTQW task which is executed in the single qubit.
Another major advantage of this method is that the whole final
states of DTQWs can be obtained, which enables one to investigate most properties of DTQWs, such as quantum correlations and coherence.


\section{Experimental Results}
In our experiments, one-particle and two-particle DTQWs with seven steps have been implemented using the one-qubit method introduced in the section II.
In this case, the total number of $a_n$ (or $b_n$) in the final state is $8$, i.e., $M=8$.
To readout the final state, we have set $\phi_k=k\phi_0$ with $\phi_0=23.6^\circ$ ($k=1, 2,..., 2M-1$).
In addition, quantum correlations and coherence in these DTQWs are systematically investigated and displayed.
The adopted measures of correlations and coherence include quantum mutual information \textit{I} \cite{q295}, entanglement \textit{E} \cite{q291}, correlated coherence $C_c$ \cite{q30}, measurement-induced disturbance $M$ \cite{q32}, and quantum coherence $C(\rho)$ from Ref.~\cite{q31}.
Note that all definitions
of the above correlations and coherence are based on the von Neumann entropy, which is
obtained from the density matrix of states, so that their behaviors could be compared with each other.
Please refer to Appendix for details of the realization of the DTQWs in linear optics, measures of the correlations and coherence, and the assisted distillation of quantum coherence. 

\subsection{Quantum correlations and coherence in one-particle DTQW}  

Photon pairs are generated by type-I spontaneous parametric down conversion in a 3-mm-thick nonlinear beta-barium borate (BBO) crystal pumped by a 100 mW diode laser (centered at 405.8 nm). The photons in one path are directly detected as the trigger and the photons in another path are injected into the setup shown in Fig.1(a) to implement a one-particle DTQW. Each step of the DTQW is achieved using a combination of two QWPs and one HWP, as shown in the Fig.1(a). We implement a seven-step DTQW using the initial state
\begin{equation}
|\phi_0\rangle=(\alpha|0\rangle_c+i\sqrt{(1-\alpha^2)}|1\rangle_c)\otimes|0\rangle ,
\end{equation}
with $\alpha\in[-1,1]$.
In this initial state, the correlations between the coin and the walker are equal to $0$ and the maxima of quantum coherence appear at $\alpha=\pm1/\sqrt{2}$.

\begin{figure*}[t]
	\begin{center}
		\includegraphics[width=14cm, clip]{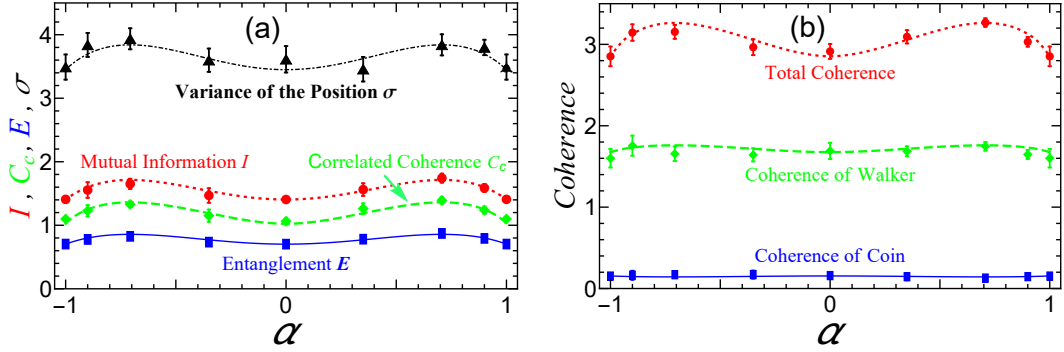}
		\caption{(a) Variance of the position $\sigma$ (black dashdotted line), quantum mutual information \textit{I} (red dotted line), correlated coherence $C_c$ (green dashed line), and entanglement \textit{E} (blue solid line) with respect to the parameter $\alpha$ of the initial state for a one-particle DTQW. (b) Effects of $\alpha$ on the quantum coherence of the coin (blue solid line), the walker (green dashed line), and the whole system (red dotted line) of a one-particle DTQW. The lines (points and error bars) represent the theoretical (experimental) results. The error bars only represent statistical errors.}\label{fig2}
	\end{center}
\end{figure*}

The whole final states of the coin and the walker are obtained from the experiment, so that the correlations and coherence of the DTQW system can be investigated. 
Appendix A2 presents the definitions of the adopted correlations and coherence measurements.

\begin{figure*}[th]
	\begin{center}
		\includegraphics[width=14cm, clip]{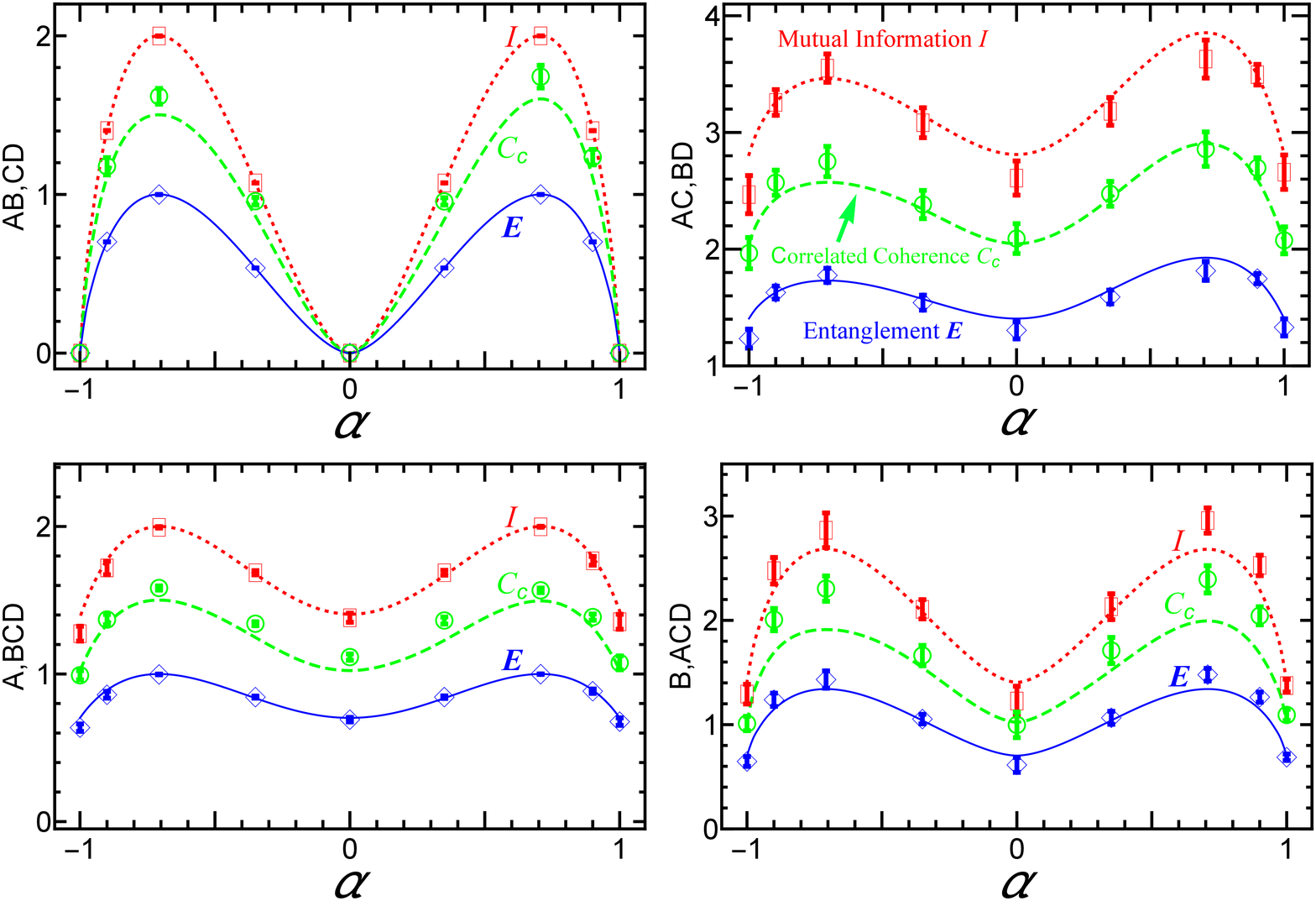}
		\caption{Quantum mutual information \textit{I} (red dotted line), correlated coherence $C_c$ (green dashed line), and entanglement \textit{E} (blue solid line) between two parts of four bipartite cases: parts AB and CD, parts AC and BD, parts A and BCD, and parts B and ACD. The coin and the walker in path 1 are denoted as A and B, while the coin and the walker in path 2 are denoted as C and D. The lines (points and error bars) represent the theoretical (experimental) results. The error bars
			only represent statistical errors. }\label{fig3}
	\end{center}
\end{figure*}

\begin{figure*}[th]
	\begin{center}
		\includegraphics[width=14 cm, clip]{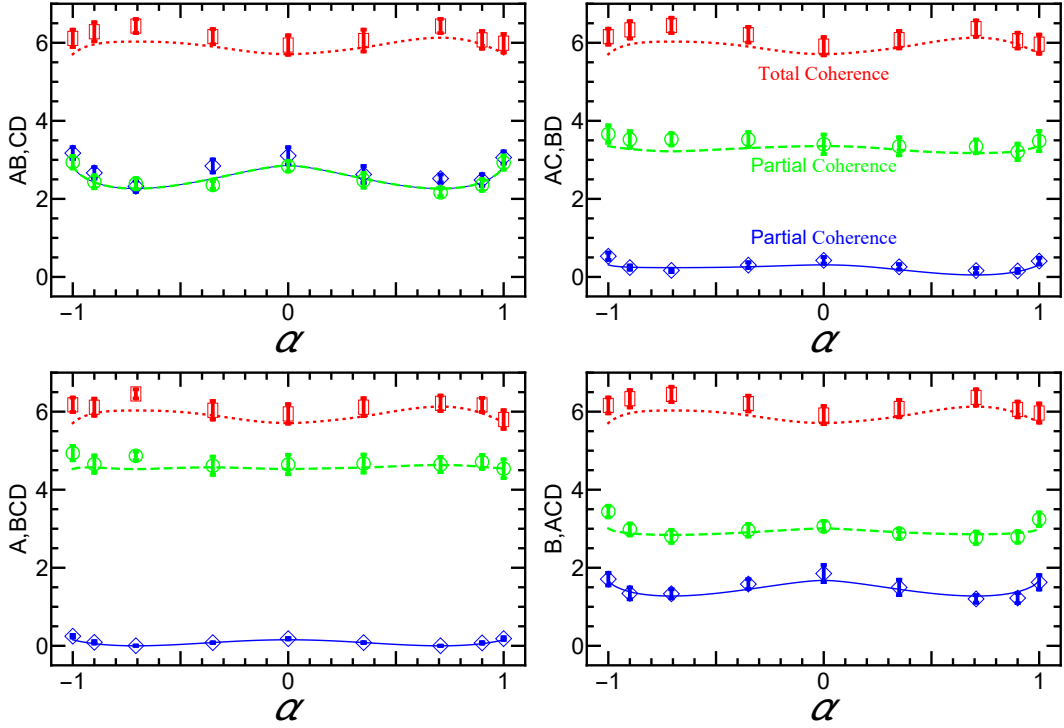}
		\caption{Quantum coherence of the whole system (red dotted line) and quantum coherence of two parts (green dashed line and blue solid line) for the four bipartite cases studied in Fig.~3. The lines (points and error bars) represent the theoretical (experimental) results. The error bars only represent statistical errors.}\label{fig4}
	\end{center}
\end{figure*}

\begin{figure*}[th]
	\begin{center}
		\includegraphics[width=16 cm, clip]{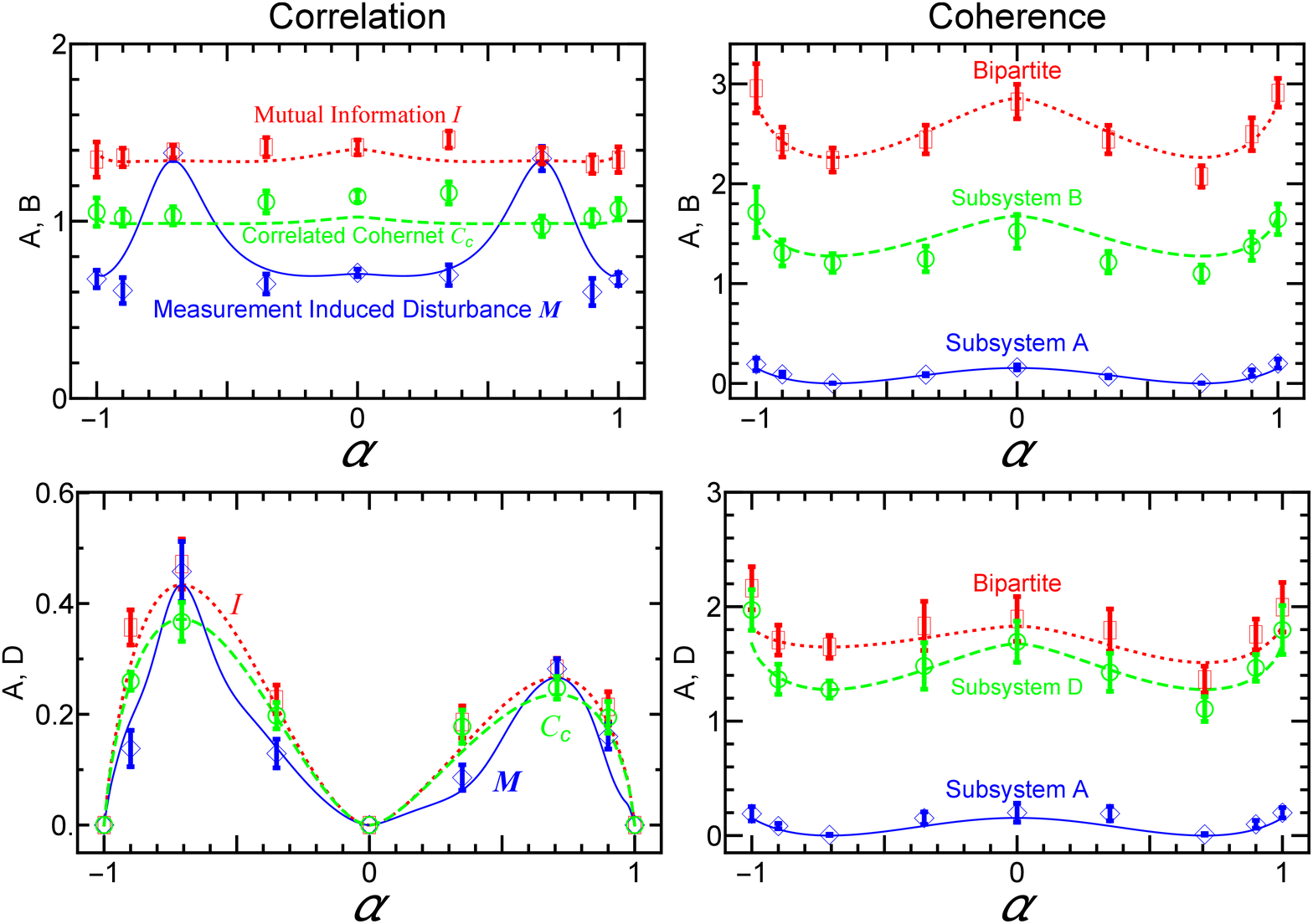}
		\caption{Left two panels: effects of $\alpha$ on the quantum mutual information \textit{I} (red dotted line), correlated coherence $C_c$ (green dashed line), and measurement-induced disturbance \textit{M} (blue solid line) between subsystems A and B, as well as between subsystems A and D, respectively. Right two panels: quantum coherence of the corresponding bipartite systems (red dotted line), and quantum coherence of  subsystem A (blue solid line) and another subsystem (green dashed line). The lines (points and error bars) represent the theoretical (experimental) results. The error bars only represent statistical errors.}\label{fig5}
	\end{center}
\end{figure*}

\begin{figure*}[th]
	\begin{center}
		\includegraphics[width=18 cm, clip]{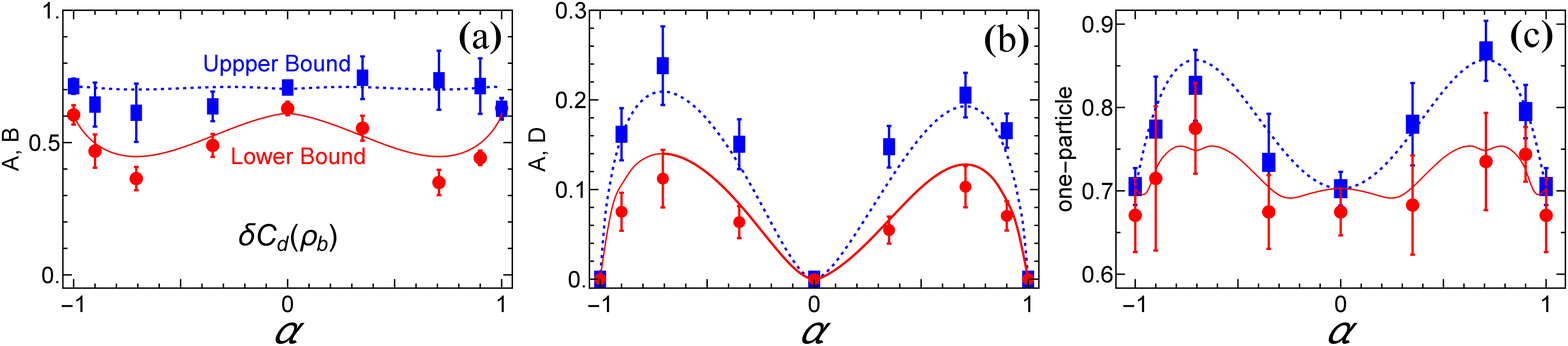}
		\caption{The upper bound of $\delta C_d(\rho_b)$ (blue dotted line) and the lower bound of $\delta C_d(\rho_b)$ (red solid line). (a) For bipartite systems consisting of subsystems A and B in the two-particle DTQW, (b) for bipartite systems consisting of subsystems A and D in the two-particle DTQW, (c) for bipartite system in the one-particle DTQW. The lines (points and error bars) represent the theoretical (experimental) results. The error bars only represent statistical errors.}\label{fig6}
	\end{center}
\end{figure*}

The effects of $\alpha$ on quantum mutual information \textit{I} \cite{q295}, entanglement \textit{E} \cite{q291}, and correlated coherence $C_c$ \cite{q30} between the coin (subsystem \textit{a}) and the walker (subsystem \textit{b}) are shown in Fig.~2(a).
Effects of $\alpha$ on quantum coherence $C(\rho)$ \cite{q31} of the coin, the
walker, and the whole system are shown in Fig.~2(b).
In addition, the variances of the walker's position $\sigma$ \cite{q301} are also obtained and shown in Fig.~2(a).

The experimental results fit well with the theoretical results.
For all $\alpha$, Fig.~2 shows that 
\begin{equation}
	I(\rho_{ab})>C_c(\rho_{ab})>E(\rho_{ab})\ \text{and}\ C(\rho_{ab}>C(\rho_{b})>C(\rho_{a}).
\end{equation}
Note that different $\alpha$ represent different initial states of the coin.
Extreme values of all correlations and coherence appear at $\alpha\sim\pm1/\sqrt{2}$, which just corresponds to the positions of maximal quantum coherence in the initial state.
It is also interesting to note that the correlations and coherence of the whole system have similar behaviors to that of the variance of the position with respect to $\alpha$. It may be inferred that both the correlations and coherence are related to the variance of the position.

\subsection{Quantum correlations and coherence in two-particle DTQW}  
Photon pairs in the entangled state
$\alpha|HV\rangle+\sqrt{(1-\alpha^2)}|VH\rangle$
can be generated from two identical down-conversion BBO
crystals \cite{q311}. This polarization-entangled state just corresponds to
the initial state of the two-particle DTQW
$(\alpha|01\rangle_c+\sqrt{(1-\alpha^2)}|10\rangle_c)\otimes|00\rangle$, in which the two coins are initially entangled.
The setup designed for a two-particle DTQW is shown in Fig.1(b).
Entangled photons are injected into path 1 and path 2,
respectively. The photons in path 1 are reflected by a beam splitter (BS) to
implement the seven-step DTQW with the optical circuit. To reuse the devices in path 1 for making the same DTQW steps, two additional HWP pairs are adopted in the path 2. In general, two optical circuits are needed if photons in two pathes are employed to realize two different DTQWs. 

For the two-particle DTQW, the whole system consists of four subsystems: two coins and two walkers.
Here we denote the coin (the walker) in path 1 as subsystem A (B) and denote the coin (the walker) in path 2 as subsystem C (D).
The two-particle DTQW system can be divided into four bipartite systems with (i) AB and CD, (ii) AC and BD,
(iii) A and BCD, (iv) B and ACD.
Due to the symmetry between two particles, these four bipartite
systems include all bipartite cases.

From the experiment on two-particle DTQW with seven steps, correlations between two parts for four bipartite cases are obtained and shown in Fig.~3,
quantum coherence of the whole system and quantum coherence of two parts are shown in Fig.~4.
For all of four bipartite cases, quantum mutual information \textit{I}, entanglement \textit{E}, and correlated coherence $C_c$ reach extremums at $\alpha\sim\pm1/\sqrt{2}$, 
where the quantum correlations between the two coins in the initial state also reach  maximum.
Likewise, quantum coherence reaches extrema at $\alpha\sim\pm1/\sqrt{2}$.
As expected, Fig.~4 indicates $I(\rho_{ab})\geq C_c(\rho_{ab})\geq E(\rho_{ab})$ for high-dimensional pure states.

The correlations between
an arbitrary coin and an arbitrary walker for the two-particle DTQW are also studied theoretically and experimentally.
For example, to obtain the correlations between
the coin in path 1 (i.e., subsystem A) and the walker in path 2 (subsystem D), we take the partial trace over
subsystems B and C and obtain a density matrix $\rho_{AD}$, which is usually a mixed state. Then we adopt quantum mutual information \textit{I}, measurement-induced disturbance $M$ \cite{q32}, and correlated coherence $C_c$ to measure the correlations between subsystem A and subsystem D.
The corresponding quantum coherence is also obtained.

The effects of $\alpha$ on the correlations between
subsystems A and B and on the correlations between subsystems A and D are obtained and shown in Fig.~5.
The corresponding coherence is also shown in Fig.~5.
All these systems consist of a coin and a walker, but behaviors of the correlations between the coin and the walker and behaviors of the coherence are very different from those for the one-particle DTQW. There is no symmetry for $\alpha\rightarrow-\alpha$ in the correlations of $\rho_{AD}$. At $\alpha\sim\pm1/\sqrt{2}$, the minima of coherence are reached, while the coherence reaches the maxima in the one-particle DTQW case. Moreover, one here can only ensure $I\geq C_c$ and $I\geq M$ for $\rho_{AD}$ and $\rho_{AB}$,   $C(\rho_{AB})>C(\rho_{B})>C(\rho_{A})$, and $C(\rho_{AD})>C(\rho_{D})>C(\rho_{A})$ as demonstrated in Fig.~5. 

\subsection{Application in assisted distillation of quantum coherence}
DTQW can generate high-dimensional multipartite states, which can be used to study how correlations and coherence change in various processes, such as in the task of assisted distillation of quantum
coherence \cite{q33,q34} in bipartite systems with high-dimensional pure or mixed quantum states. In this task, two parties, Alice (with subsystem \textit{a}) and Bob (with subsystem \textit{b}), are involved. 
Their goal is to maximize the quantum coherence of Bob's subsystem \textit{b} by Alice performing arbitrary quantum operations on subsystem \textit{a}, while Bob is restricted to just incoherent operations on his subsystem \textit{b}. The classical communication between Alice and Bob is allowed. This is referred to as local quantum-incoherent operations and classical communication (LQICC). 

The definition of maximal increase in distillable coherence of subsystem \textit{b} [$\delta C_d(\rho_b)$] is introduced in Appendix A3.
The $\delta C_d(\rho_b)$ for pure states can be easily calculated, while it is challenging to obtain the $\delta C_d(\rho_b)$ for mixed states.
We introduce an upper bound and a lower bound of $\delta C_d(\rho_b)$ for mixed states in Appendix A3. With the bounds, some properties of $\delta C_d(\rho_b)$  for mixed states can be obtained and studied.

Two bounds of $\delta C_d(\rho_b)$ with different values of $\alpha$ for mixed states $\rho_{AB}$ and $\rho_{AD}$ in the seven-step two-particle DTQW are investigated both theoretically and experimentally, as shown in
Fig.~6(a) and Fig.~6(b). Here, both bipartite systems consist of a
two-dimensional subsystem \textit{a} (the coin) and a high-dimensional
subsystem \textit{b} (the walker). 
Experimental results fit well with the theoretical results, and the lower bound of $\delta C_d(\rho_b)$ is under and close to its upper bound. Obviously, $\delta C_d(\rho_b)$ is confined to the region between the
upper and lower boundaries. By comparing these results with the results of $\rho_{AB}$ and $\rho_{AD}$ in Fig.~5, it may be inferred that $\delta C_d(\rho_b)$ is related to the correlations between subsystem \textit{a} and subsystem \textit{b}. Moreover, one has $\delta C_d(\rho_b)\leq C_c(\rho_{ab})\leq I(\rho_{ab})$ for all $\alpha$.

To verify the validity of the bounds of $\delta C_d(\rho_b)$, we also study $\delta C_d(\rho_b)$ for the pure state of the seven-step one-particle DTQW in the experiment, as shown in Fig.~6(c). For a pure state, the upper bound is equal to $\delta C_d(\rho_b)$.  Figure~6(c) shows that the lower bound is indeed lower than $\delta C_d(\rho_b)$ for pure states of the one-particle DTQW.

\section{Discussion and Conclusion} 
In our experiments, the standard DTQWs with seven steps are performed in linear optical setups, from which all final joint states of the walker and the coin for the seven-step DTQWs are obtained. Based on the joint states from the experiments, the quantum correlations and coherence in the seven-step DTQWs are investigated systematically. As all DTQW operations are implemented in the two-dimensional polarization space of a single photon, a seven-step DTQW is realized using seven combinations of wave-plates, which is very economic and efficient. Moreover, the single-photon measurement after seven-step DTQWs is extremely simple.
 
To obtain the final joint state of the walker and the coin for a DTQW after $N$ steps, it appears that our DTQW experiment must be performed ($2N+1$) times with different $\phi_k$, in order to achieve ($2N+1$) equations of Eq.~(\ref{e}). Actually, in other
DTQW experiments, which use orthogonal states and thus require many more qubits  \cite{q9,q130,q13,q14}, 
an $N$-step DTQW has to be implemented $2(N+1)$ times to measure $2(N+1)$ probabilities $|a_n|^2$ and $|b_n|^2$.
Moreover, it is challenging to obtain the complex numbers $a_n$ and $b_n$ in those DTQW experiments.

If a DTQW experiment is implemented in a single qubit, there is a limitation on the number of DTQW steps, which mainly depends on the accuracy of the state measurement. 
A key criterion is that the non-orthogonal states, which encode the states of the walker and the coin, must be distinguishable in the measurement.
As shown in our experiment, by using a single qubit at least seven DTQW steps can be realized in the linear optical setups.
The number of performable DTQW steps can be increased if the experimental accuracy is improved, such as adopting devices of high precision.
Another efficient way of increasing the number of performable DTQW steps is extending the one-qubit method to more qubits and encoding the DTQW states by non-orthogonal states in the joint space of the qubits. 
In this case, DTQWs with a large number of steps could be performed even if an extremely small number of qubits are used.

In summary, we have proposed a novel one-qubit DTQW implementation method, which makes it easy to obtain the final joint states of DTQWs and thus enables one to investigate most properties of one-particle and multi-particle DTQWs.
Based on this method, we have experimentally implemented one-particle and two-particle DTQW with seven steps in linear optics. By obtaining the final joint states of the DTQWs, we have investigated quantum correlations and coherence in one-particle and two-particle DTQWs experimentally.
Furthermore, we have shown how to use the two-particle DTQW, which can generate high-dimensional states of multipartite systems, to study the correlations and coherence rules, i.e., the assisted distillation of quantum coherence for high-dimensional mixed states.

Since only single-qubit operations are required,
the proposed one-qubit approach for realizing DTQWs is quite general, which can be applied to implement single-qubit-based DTQW experiments in various physical systems (using much less resources), such as linear optics, circuit QED and trapped ions. This work paves an avenue to study properties, such as correlations and coherence, of the DTQWs and may extend their applications. The present work also opens a new direction to study quantum correlation and coherence rules of high-dimensional states in experiments. 
Furthermore, the proposed one-qubit encoding method is expected to have applications in quantum algorithms and quantum simulations.\\
\\
\textbf{Acknowledgment} This work is supported in part by the National Natural
Science Foundation of China under Grants No.11504075,
No.11774076, No.11974096, No.11775065, No.11074062,
No.11374083, and No.U21A20436,
and the Key-Area Research and Development Program of Guangdong
Province (2018B030326001).
F.N. is supported in part by:
Nippon Telegraph and Telephone Corporation (NTT) Research,
the Japan Science and Technology Agency (JST) [via
the Quantum Leap Flagship Program (Q-LEAP),
the Moonshot R\&D Grant Number JPMJMS2061, and
the Centers of Research Excellence in Science and Technology (CREST) Grant No. JPMJCR1676],
the Japan Society for the Promotion of Science (JSPS)
[via the Grants-in-Aid for Scientific Research (KAKENHI) Grant No. JP20H00134 and the
JSPS–RFBR Grant No. JPJSBP120194828],
the Army Research Office (ARO) (Grant No. W911NF-18-1-0358),
the Asian Office of Aerospace Research and Development (AOARD) (via Grant No. FA2386-20-1-4069), and
the Foundational Questions Institute Fund (FQXi) via Grant No. FQXi-IAF19-06.\\
\\
\textbf{Conflicts of interest}\ \
The authors declare no conflicts of interest.


\setcounter{equation}{0}
\renewcommand{\theequation}{a\arabic{equation}}

\section*{Appendix}
\textbf{A1 \ \ The Implementation of DTQWs in linear optics}

The coin state is encoded by
the horizontal polarization state $|H\rangle$ and the vertical polarization state $|V\rangle$ of a single photon, and the walker's positions
are encoded by the phases of $|H\rangle$ and $|V\rangle$ of the single photon.
The general state of Eq.~(3) in the main text is encoded as
\begin{equation}
	\frac{1}{N}\sum_n(a_ne^{-in\phi}|H\rangle+b_ne^{in\phi}|V\rangle).
\end{equation}
Then the initial state of the one-particle DTQW in our experiment is encoded as \begin{equation}
	\psi_0=(\alpha|0\rangle_c+i\sqrt{1-\alpha^2}|1\rangle_c)|0\rangle\rightarrow\alpha|H\rangle+i\sqrt{1-\alpha^2}|V\rangle,
\end{equation}
which is easily prepared by injecting a horizontally polarized photon into a HWP and a QWP, as shown in Fig.~1(a) of the main text.
And the initial state of the two-particle DTQW is encoded as
\begin{equation}
	(\alpha|01\rangle_c+\sqrt{(1-\alpha^2)}|10\rangle_c)\otimes|00\rangle\rightarrow \alpha|HV\rangle+\sqrt{1-\alpha^2}|VH\rangle,
\end{equation}
which is generated using two identical BBO crystals.

Next, we introduce how to realize the DTQW operator $U=T\cdot S$ in a linear optical system. In the two-dimensional polarization space, the coin operator $S$ can be expressed as
\begin{equation}
	S(\theta)=\left(
	\begin{array}{cc}
		\cos\theta & -\sin\theta  \\
		\sin\theta  & \cos\theta%
	\end{array}%
	\right)=e^{-i\theta\sigma_y}, 
\end{equation}
and the shift operator $T$ is
\begin{equation}
	T=\left(
	\begin{array}{cc}
		e^{-i\phi} & 0  \\
		0  & e^{i\phi}%
	\end{array}%
	\right)=e^{-i\phi\sigma_z}, 
\end{equation}
where $\sigma_y$ and $\sigma_z$ are Pauli matrices.
In our DTQW experiments, $\theta=\pi/4$ is adopted.
The unitary operator $U=e^{-i\phi\sigma_z}e^{-i\theta\sigma_y}$ is realized by a combination of two QWPs and one HWP:
\begin{equation}
	U=U_{\textrm{QWP}}\left({\pi\over4}\right)U_{\textrm{HWP}}\left({\phi\over2}-{\theta\over2}+{\pi\over4}\right)U_{\textrm{QWP}}\left({\pi\over4}+\theta\right).
\end{equation}
As shown in Fig.~1 of the main text, an $n$-step DTQW needs $n$ combinations of waveplates. 
After an \textit{n}-step DTQW implementation, the state of single photons can be easily obtained via tomography measurements.
Then, the final state of the DTQW after $n$ steps can be readout using the encoding method.
\\

\textbf{A2 \ \ Measures of correlations and coherence}

The coherence in the DTQW system and its subsystems will be studied using the relative entropy of coherence \cite{q31}. 
Correlation between arbitrary two subsystems of DTQW will
be studied using quantum mutual information \cite{q295}, measurement-induced
disturbance \cite{q32}, entanglement \cite{q291}, and correlated coherence \cite{q30}. 
The correlated coherence is just the quantum coherence contained within correlation, which can be used as a kind of quantum correlation. All definitions of the above coherence and correlations are based on the von Neumann entropy, which is obtained from the density matrix of states, so that their behaviors are comparable. Consider a bipartite state
$\rho_{ab}$ of subsystem \textit{a} and subsystem \textit{b}. The measure of correlations and coherence are defined as follows.

1. Quantum mutual information $I$ is usually used to quantify the total correlation
between subsystem $a$ and subsystem $b$, with
\begin{equation}
	I(\rho_{ab})=S(\rho_a)+S(\rho_b)-S(\rho_{ab}), 
\end{equation}
where $\rho_a=\text{tr}_b\ \rho_{ab}$, $\rho_b=\text{tr}_a\ \rho_{ab}$, and $S(\rho)=-\text{tr}[\rho \log_2\rho]$ denoting the von Neumann entropy.

2. The measurement-induced disturbance $M$ quantifies a sort of
quantum correlation between two subsystems, with
\begin{equation}
	M(\rho_{ab})=I(\rho_{ab})-I(\Pi(\rho_{ab})), 
\end{equation}
where $\Pi(\rho_{ab})=\sum_{ij}\Pi_a^i\otimes\Pi_b^j\ \rho \Pi_a^i\otimes\Pi_b^j$ is a classical state;
and the complete projective measurements $\{\Pi_a^i\}$ and $\{\Pi_b^j\}$ are
induced by the spectral resolutions of $\rho_a=\sum_i p_a^i\Pi_a^i$
and $\rho_b=\sum_j p_b^j\Pi_b^j$, respectively. Here $M(\rho_{ab})$
represents the disturbance of the correlation (i.e., quantum mutual
information) induced by the measurement $\Pi$.
Note that the chosen measurement $\Pi$ leaves $\rho_a$, $\rho_b$, and the marginal
information invariant, and $M(\rho)$ is invariant under local
unitary transformations. Moreover, for pure states, $M$ coincides with the conventional entanglement \textit{E}, i.e., $M(\rho_{ab})=S(\rho_a)=S(\rho_b)$.

3. The entanglement \textit{E} for a pure state of bipartite system can be quantified by the von
Neumann entropy, i.e.,
\begin{equation}
	E(\rho_{ab})=S(\rho_a)=S(\rho_b).
\end{equation}
For high-dimensional mixed states, measure of entanglement will not be
illustrated, because there is still no suitable definition of entanglement.

4. The relative entropy of coherence $C$ is calculated as
\begin{equation}
	C(\rho)=S(\rho^D)-S(\rho), 
\end{equation}
where $\rho^D$ is the diagonal version of $\rho$, which only retains
the diagonal elements of $\rho$.

5. The correlated coherence $C_c$ is just the local coherence subtracted
from the total coherence:
\begin{equation}
	C_c(\rho_{ab})=C(\rho_{ab})-C(\rho_{a})-C(\rho_{b}).
\end{equation}
This correlated coherence can be rewritten as
\begin{equation}
	\begin{aligned}
		C_c&=[S(\rho_{a})+S(\rho_{b})-S(\rho_{ab})]-[S(\rho_{a}^D)+S(\rho_{b}^D)-S(\rho_{ab}^D)]  \\
		&=I(\rho_{ab})-I(\rho_{ab}^D),
	\end{aligned}
\end{equation}
which is similar to the definition of measurement-induced disturbance
$M$. Though the correlated coherence is derived from the quantum
coherence, it can be treated as a sort of measurement-induced
disturbance of correlation under a simple measurement based on the basis of $\rho_{ab}$.

In general, one has the relation $I(\rho_{ab})\geq C_c(\rho_{ab})\geq E(\rho_{ab})$, because $I(\rho_{ab})$ represents the quantum and classical correlation, $C_c(\rho_{ab})$ represents the quantum correlation, and $E(\rho_{ab})$ represents the quantum entanglement, which can be treated as a minimal definition of quantum correlation.
\\

\textbf{A3 \ \ Maximal increase in distillable coherence}

It has been proved  \cite{q33} that for a pure state $|\phi\rangle_{ab}$ of the two subsystems \textit{a} and \textit{b}, the maximal increase in
distillable coherence of subsystem \textit{b} is equal to the von Neumann entropy of $\rho_b$, i.e., 
\begin{equation}
	\delta C_d(\rho_b)\equiv C_d^{a|b}(|\phi\rangle_{ab})-C_d(\rho_b)=S(\rho_b)
	,\label{delta}
\end{equation}
where $C_d^{a|b}(|\phi\rangle_{ab})$ is the distillable coherence of
collaboration and $C_d(\rho_b)=C(\rho_b)\equiv S(\rho^D_b)-S(\rho_b)$
is the distillable coherence of
$\rho_b$. For four bipartite cases shown in Fig.~3 and Fig.~4 of the main text, 
the corresponding $\delta C_d(\rho_b)$, $C_d^{a|b}(|\phi\rangle_{ab})$ and $C_d(\rho_b)$ can be easily obtained and used to demonstrate the relationship in Eq.~(13) of the main text.

It is more interesting to study the assisted distillation of quantum coherence with mixed states. For a general mixed state $\rho_{ab}$, there is no deterministic equation of $\delta C_d(\rho_b)$ like the Eq.~(13) of the main text
because of the difficulty for calculating $C_d^{a|b}(\rho_{ab})$.
Instead, the bounds of $C_d^{a|b}(\rho_{ab})$ can be obtained and used to study the properties of $\delta C_d(\rho_b)$ for mixed states.
In the following, we introduce an upper bound and a lower bound of $\delta C_d(\rho_b)$.

1.\! The distillable coherence of collaboration is bounded \cite{q33} according to
\begin{equation}
	C_d^{a|b}(\rho_{ab})\leq C_r^{a|b}(\rho_{ab}),
\end{equation}
where $C_r^{a|b}(\rho_{ab})=S[\Delta_b(\rho_{ab})]-S(\rho_{ab})$
with $\Delta_b(\rho_{ab})=\sum_i(I_a\otimes|i\rangle_b\langle i|)\rho_{ab}(I_a\otimes|i\rangle_b\langle i|)$, which is
defined for a fixed incoherent basis $\{|i\rangle_b\}$.
Thus, one has an upper bound for $\delta C_d(\rho_b)$:
\begin{equation}
	\delta C_d(\rho_b)\leq C_r^{a|b}(\rho_{ab})-C_d(\rho_b),\label{up}
\end{equation}
which is calculated from $\rho_{ab}$.

2. There is no simple lower bound for $C_d^{a|b}(\rho_{ab})$. To obtain a lower bound, one can implement a range of special LQICC operations on the bipartite system and then treat the maximal coherence obtained from subsystem \textit{b} as a lower bound of $C_d^{a|b}(\rho_{ab})$. In order to obtain this lower bound in experiments, von Neumann measurements of subsystem \textit{a} are adopted as LQICC operations, which will be performed on one copy of the state, instead of complex collective operations on many copies. In this way, one obtains a lower bound for the distillable coherence of collaboration of one-copy, which is also a lower bound for $C_d^{a|b}(\rho_{ab})$.  We denote this
lower bound of $C_d^{a|b}(\rho_{ab})$ as $C_l(\rho_b)$ and obtain the lower bound of $\delta C_d(\rho_b)$:
\begin{equation}
	\delta C_d(\rho_b)\geq C_l(\rho_b)-C_d(\rho_b).\label{low}
\end{equation}	

To obtain $C_l(\rho_b)$, we have performed von
Neumann measurements on subsystem \textit{a} in the basis $\cos(\theta)|0\rangle+e^{i\phi}\sin(\theta)|1\rangle$
and $e^{-i\phi}\sin(\theta)|0\rangle-\cos(\theta)|1\rangle$. Then the maxima of $C_d(\rho_b)$ are found as $C_l(\rho_b)$ for each $\alpha$ with $\theta\in[0,\pi]$ and
$\phi\in[0,2\pi]$.

\end{document}